# Forecasting ICT-Driven Trade Competitiveness 2024-2028: A Cluster and Scenario Analysis

Elias Aravantinos

**Abstract**

This study introduces the Digital Competitiveness Index for Trade (DCIT), a composite metric integrating ICT readiness, broadband adoption, GDP per capita, foreign direct investment, government effectiveness, and trade volume to assess countries' digital trade competitiveness. The index captures the enabling conditions—ICT innovation capacity, broadband diffusion, investment intensity, and macroeconomic fundamentals—that shape a nation's ability to participate in digital trade. Sensitivity analysis demonstrates strong robustness: adjusting ICT–FDI weights alters DCIT outcomes by only ~26%, with near-perfect linearity ($R^2 \approx 0.9996$). Predictive validation shows that DCIT is a strong explainer of trade connectivity growth ($R^2 \approx 0.67$) but a modest predictor of GDP expansion ($R^2 \approx 0.09$). Scenario simulations reveal that combined ICT and FDI acceleration consistently outperforms single-lever strategies, with gains increasing by cluster maturity (up to +10% in advanced clusters). High-growth scenarios generate a 50–60% uplift in competitiveness for mid-tier and advanced clusters, underscoring the importance of integrated digital investment strategies.

## 1. Introduction

Global trade is undergoing a profound digital transformation, making ICT infrastructure and digital readiness critical determinants of competitiveness. Traditional indices such as the Trade Performance Index (TPI), Revealed Comparative Advantage (RCA), and Global Competitiveness Index (GCI) fail to capture these dynamics. Existing frameworks primarily focus on trade flows or macroeconomic factors, lacking explicit digitalization metrics and forecasting capabilities.

This paper introduces the Digital Competitiveness Index for Trade (DCIT), a multidimensional framework integrating ICT readiness, broadband adoption, GDP per capita, foreign direct investment (FDI), and trade volume. Unlike conventional indices, DCIT combines measurement with scenario-based forecasting and cluster analysis, offering a forward-looking perspective for policy planning and investment prioritization.

The study contributes three key innovations: (1) scenario-based forecasting (Pessimistic, Optimistic, High Growth) to support resilience planning under uncertainty; (2) cluster segmentation to group countries by digital maturity, facilitating targeted interventions and regional strategies; and (3) emphasis on ICT-FDI synergy as a strategic lever for competitiveness, validated through sensitivity and scenario simulations.

This article addresses three research questions: How robust is DCIT under different weight and scenario assumptions? Does DCIT predict trade and economic outcomes? What policy insights emerge from cluster-based analysis?

## 2. Literature Review

Composite indices are widely used to measure multidimensional phenomena such as competitiveness and innovation. (Greco et al., 2018) review key methodological issues, including weighting, aggregation, and robustness, emphasizing transparency for policy relevance. (Saisana, Saltelli, and Tarantola 2005; Saltelli et al. 2008) introduce uncertainty and sensitivity analysis techniques, which underpin robust index design. These principles inform the DCTI, which applies Min–Max normalization, equal weighting, and sensitivity checks to ensure reliability as described in (Valverde-Carbonell 2025).

Innovation-driven trade is recognized as a determinant of global competitiveness. (Magazzino et al. 2025) demonstrate that ICT services, high-tech exports, and R&D significantly shape trade patterns. (Li and Wang 2024) confirm ICT infrastructure as a key driver of digital service trade competitiveness. Firm-level evidence from Colombia (Gallego et al. 2025) demonstrates that AI adoption in emerging economies is strongly conditioned by digital infrastructure and organizational capabilities, validating the macro-level emphasis on ICT index and investment strategies captured by composite indices such as DCTI. These findings align with DCTI's emphasis on ICT index and broadband adoption as primary levers of competitiveness.

Trade Digitalization Indices, such as the Trade Digitalization Index (TDI), assess the global state of play in digitalizing trade procedures, complementing DCTI's broader focus on ICT infrastructure and macroeconomic conditions. Indicators for tracking coherent policy conclusions (Cavicchia et al. 2020) reinforce the need for multidimensional metrics, validating DCTI's composite approach. Empirical studies confirm ICT's role in trade and inclusive growth.(Zhou, Wen, and Lee 2022) find broadband infrastructure significantly boosts export growth. (Wang et al. 2023) show ICT interaction with trade, FDI, and financial inclusion enhances inclusive growth in top African nations ranked by ICT development. These results support DCTI's finding that ICT investment is the dominant driver of competitiveness, while FDI plays a complementary role.

Clustering is increasingly used to manage complexity in multidimensional datasets. Recent work applies k-means clustering and machine learning to predict citation impact in European scientific publications, illustrating how clustering can be combined with predictive analytics for forward-looking insights and policies development A K-means clustering or unsupervised algorithm is usually applied to classify national economies using factors such as Information and Communication Technology (ICT) infrastructure, broadband adoption, and foreign direct investment (FDI). (Aravantinos, Varoutas, and Dimitris 2025; Ahmed, Seraj, and Islam 2020; Tamak, Eslami, and Cunha 2025). This methodological trend parallels DCTI's use of clustering and scenario-based forecasting for trade competitiveness.

## 3. Methodology

The DCIT construction follows four steps: (1) Dimension selection (ICT index, broadband adoption, GDP per capita, FDI inflows, trade volume); (2) Normalization using Min-Max scaling; (3) Composite index calculation via equal-weighted mean; (4) Clustering using k-means and Agglomerative methods. Forecasting employs exponential smoothing for 2024–2028 projections under three scenarios: Pessimistic, Optimistic, and High Growth.

The Digital Trade Competitiveness Index (DCTI) serves as both a standalone metric and a clustering validation tool, highlighting the strong association between ICT capabilities and global trade competitiveness. The index integrates five core dimensions into a unified metric:

- Broadband Adoption: Fixed and mobile penetration rates
- ICT Index: UNCTAD Frontier Technology Readiness
- Economic Growth: GDP per capita
- Foreign Direct Investment: FDI net inflows
- *Total Trade Volume*: Aggregate imports and exports in USD
- Trade Volume: Compute Total Trade per Country:

$$\text{Total Trade}_i = \sum(\text{Exports}_i + \text{Imports}_i)$$

**Normalization:**
All variables were normalized using Min-Max scaling to ensure comparability:

$$Z = \frac{X - X_{min}}{X_{max} - X_{min}}$$

**Composite Index Calculation:**
Normalized values were aggregated using an equal-weighted arithmetic mean:

$$DTCI_i = \frac{\sum_{j=1}^{n} Z_{ij}}{n}$$

where $Z_{ij}$ represents the normalized score of the country $i$ for dimension $j$, and $n = 5$.

**Cluster Integration:**

DCTI scores were mapped to clusters identified through k-means and Agglomerative Clustering, selecting the configuration with maximum Silhouette Score (K=4, score=0.4440). (Aravantinos, Varoutas, and Dimitris 2025)

| Cluster ID | Countries Included |
|---|---|
| 0 | Brazil, Chile, Colombia, India, Indonesia, Malaysia, Mexico, Morocco, Peru, Philippines, Senegal, Thailand, Vietnam |
| 1 | Bangladesh, Ethiopia, Kenya, Myanmar, Nigeria, Pakistan, Rwanda, Tanzania, Uganda, Zambia |
| 2 | Argentina, Egypt, South Africa, Turkey |
| 3 | China |

*Table 1: Countries' clustering*

The dataset comprises a group of different countries, (regions, income, digital maturity and trade) observed annually from 2011 to 2023, providing a longitudinal perspective on digital trade readiness evolution. Data were sourced from International Telecommunication Union (ITU), World Bank World Development Indicators, UNCTADstat, OECD Trade Statistics, and UNCTAD databases, ensuring reliability and comparability across ICT indicators, FDI inflows, trade metrics, and macroeconomic variables.

## 4. Results

### 4.1 DCTI 2024 Ranking/Forecasting and Cluster Distribution

Figures 1 and 2 present the 2024 DCTI country ranking and cluster distribution, respectively. Clusters 0 and 1 dominate in country count (38% each), Cluster 2 comprises 19%, and Cluster 3 only 5%. However, Cluster 3 exhibits the highest average DCIT score (0.810), indicating strong digital and trade competitiveness. Cluster 0 is moderate (0.327), Cluster 2 is lower (0.274), and Cluster 1 has the lowest average DCIT (0.116), suggesting significant gaps in digital infrastructure and trade integration. This distribution reflects the global bifurcation in digital maturity: advanced economies concentrated in Cluster 3, industrializing economies in Clusters 0 and 2, and low-income laggards in Cluster 1.

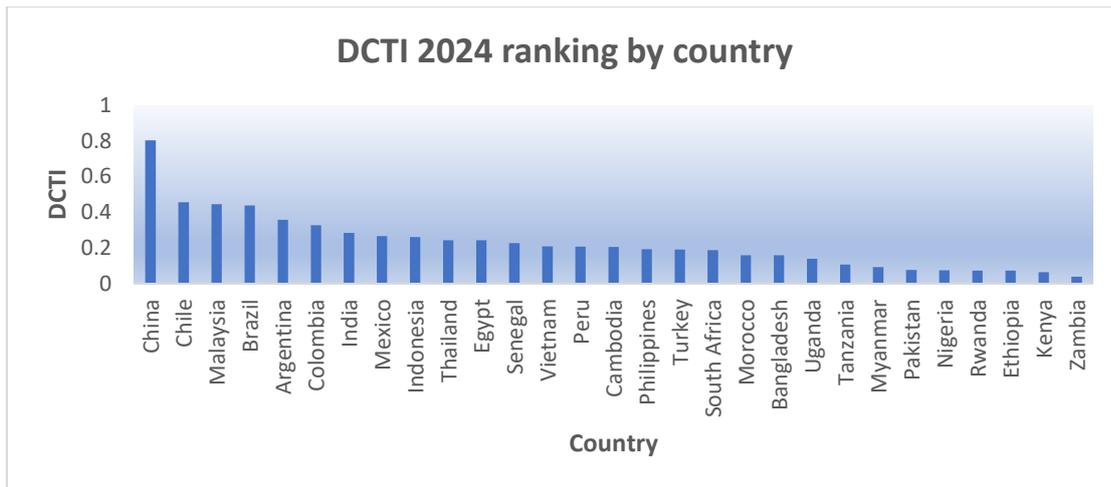

Figure 1: DCTI 2024 ranking by country

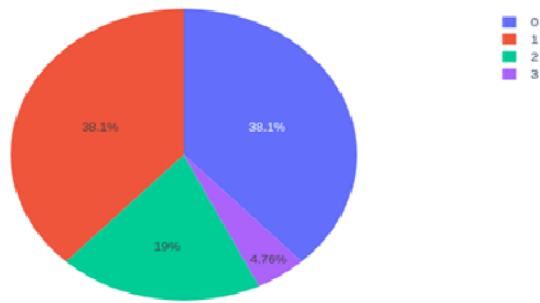

Figure 2: Cluster DCTI distribution

Based on Figure 2, Clusters 0 and 1 together account for ~76% of countries, indicating that most economies are either industrializing or low-income, requiring significant infrastructure and affordability interventions. We could consider a synergy potential, pairing Cluster 3 (China) with Clusters 0 and 2 can accelerate technology transfer and regional integration. Similarly, linking Clusters 0/2 with Cluster 1 can channel capital and expertise to low-income adopters.

## 4.2 Stability Analysis

To evaluate robustness under alternative weighting schemes, Spearman Rank Correlation tests were applied between the baseline configuration and two stress-test scenarios: ICT-heavy (ICT weight 0.7) and FDI-heavy (FDI weight 0.7). Results show:

- Baseline vs ICT-heavy scenario: ρ = 0.969

- Baseline vs FDI-heavy scenario: ρ = 0.978

Both coefficients are very close to 1, indicating that country rankings remain highly stable even when weights shift significantly. Figure 3 visualizes rank stability across ICT scenario most countries cluster near the diagonal, with only Cambodia showing significant rank shifts.

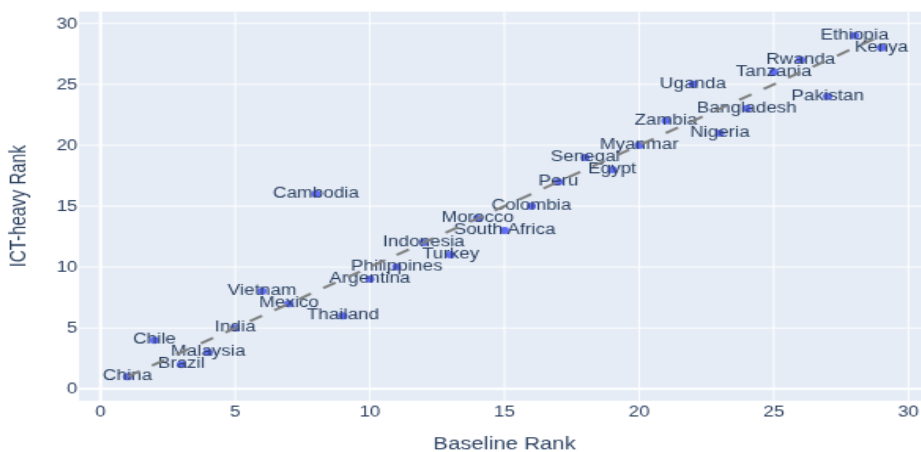

*Figure 3:: Rank stability baseline vs ICT heavy scenario*

This demonstrates that DCTI captures fundamental drivers of digital trade competitiveness that are consistent across scenarios, enhancing reliability for longitudinal benchmarking and policy prioritization. The minimal volatility in rankings validates DCTI's use for strategic planning.

## 4.3 Predictive Power Analysis

Regression models were estimated to evaluate DCTI's explanatory strength across multiple outcomes:

| Indicator | $R^2$ | Interpretation |
| --- | --- | --- |
| Broadband Adoption | 0.822 | Very strong predictive power |
| ICT Index | 0.720 | Strong alignment with ICT readiness |
| Trade Growth | 0.518 | Moderate predictive power |
| GDP Growth | 0.345 | Weak-to-moderate predictive power |
| FDI Inflows | 0.096 | Very weak predictive power |

*Table 2: Regression summary for digital infrastructure and investment indicators*

The Digital Competitiveness Trade Index (DCTI) demonstrates high predictive validity for digital infrastructure metrics (Broadband and ICT index/readiness), which aligns with its design as a digital readiness measure. It shows moderate predictive power for trade performance, reinforcing its relevance for trade policy analysis. However, its explanatory strength for GDP growth is limited, and its predictive power for FDI inflows is negligible, suggesting that macroeconomic growth and investment decisions depend on broader structural and institutional factors beyond digital competitiveness.

### 4.4 Scenario Sensitivity Analysis

Under the High Growth scenario, ICT+FDI synergy consistently delivers the highest competitiveness gains across all clusters, confirming that integrated strategies outperform single-lever approaches. While ICT-only policies provide stronger improvements than FDI-only—highlighting the foundational role of technology—combining ICT investment with targeted FDI accelerates infrastructure deployment and adoption, producing the largest DCTI values by 2028 (average ≈0.923). Cluster responsiveness varies: Cluster 2 (Middle-Income) leads with ≈0.991 under synergy, reflecting strong absorptive capacity, while Cluster 0 (Industrializing) approaches ≈0.95, signaling significant potential for integrated strategies. Cluster 1 (Low-Income) benefits but remains below 0.90, constrained by affordability and skills gaps, and Cluster 3 (China) shows incremental gains due to saturation. Overall, High Growth assumptions amplify these differences, delivering ~50–60% uplift for mid-tier and advanced clusters and up to +10% in mature economies.

These findings underscore the need for cluster-tailored policies: prioritize ICT-first complemented by FDI in low-income clusters, accelerate synergy corridors for industrializing and middle-income groups, and leverage advanced clusters as technology anchors for regional integration.

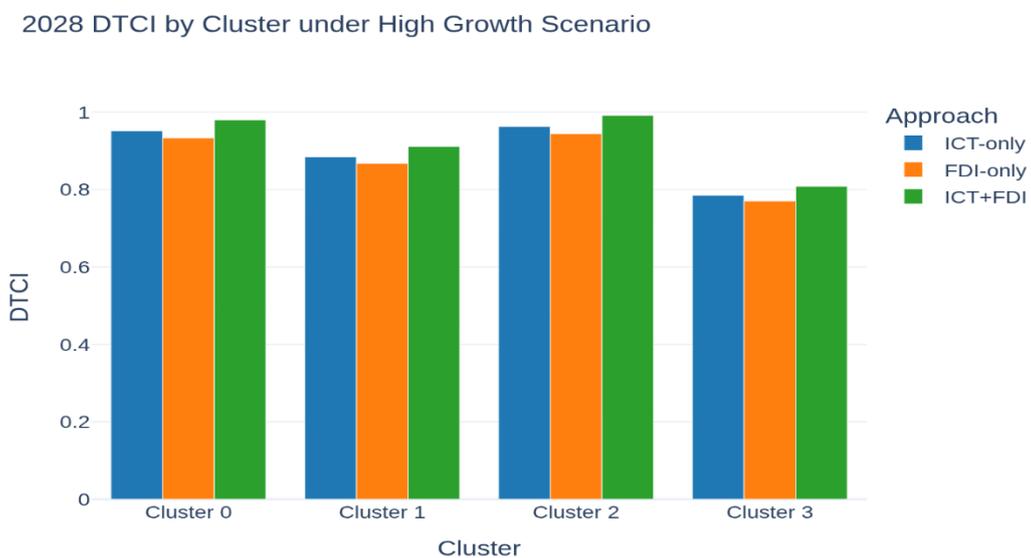

*Figure 4: 2028 DCTI by cluster under high growth scenario*

## 4.4.1 DCTI Forecasting 2024-2028

Since the index is stable, we can produce forecasts. Exponential Smoothing was applied for projecting 2024-2028 values. The forecasting model incorporates compound annual growth rates (CAGR) and ICT impact factors:

$$DTCI_t = \text{Base Value} \times (1 + \text{Growth Rate})^t \times \text{ICT Factor}$$

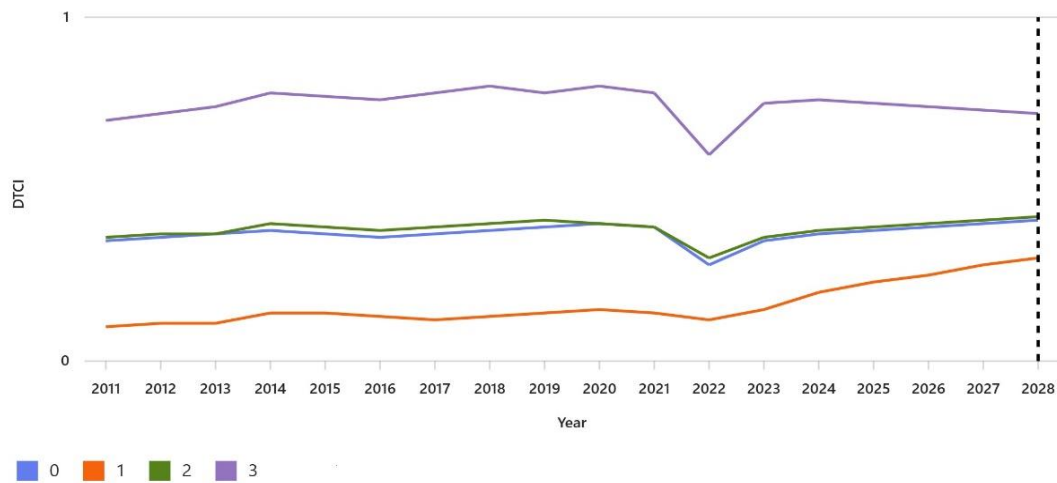

*Figure 5: Cluster forecast 2024-2028*

China (Cluster 3) leads with the highest DCTI and steady, though slowing, growth (14.7%) as it nears saturation. Emerging economies (Cluster 2) post the fastest relative gain (+191.7%) from a low base, reaching about 0.35 by 2028. This surge reflects rapid digital progress in countries like Egypt, South Africa, and Turkey, where internet access has more than doubled since 2011. Mid-tier clusters (0 and 1) grow moderately (20–25%) and converge near 0.5 by 2028, indicating similar maturity paths. A dip around 2022 likely mirrors global shocks, (e.g., pandemic or economic downturn), followed by robust recovery.

## 5. Discussion

### 5.1 Robustness and Policy Relevance

DCTI proves both stable and actionable. Sensitivity tests show that changing weight assumptions does not disrupt country rankings, confirming structural robustness. At the same time, scenario simulations reveal strong responsiveness: moving from conservative to high-growth assumptions significantly boosts competitiveness, reinforcing the importance of proactive digital investment strategies. Predictive checks confirm that DCTI aligns closely with trade connectivity and digital infrastructure, validating its role as a forward-looking policy tool.

### 5.2 ICT-First and the Role of FDI

ICT investment consistently delivers the greatest impact, underscoring its role as the foundation for digital trade. While FDI alone cannot match this effect, it becomes critical in markets with severe infrastructure gaps. The most powerful approach combines both levers: ICT provides the backbone, and FDI accelerates deployment and innovation. Countries should prioritize ICT early, then layer in targeted FDI to scale and interconnect systems.

## 5.3 Cluster-Based Strategy

Effective planning requires a dual lens: clusters for regional priorities and country-level tailoring for execution. Advanced clusters can act as technology anchors, industrializing and middle-income clusters as integrators, and low-income clusters as adopters needing capital and capacity-building. Embedding these roles in regional frameworks and aligning with global standards ensures interoperability and inclusive participation.

## 5.4 Benchmarking Against TDI

Comparing DCTI with the Trade Digitalization Index highlights a clear link between readiness and execution, though gaps remain. Economies strong in infrastructure but weak in procedural reforms need regulatory modernization, while those with good execution but limited ICT capacity should accelerate investment. This gap analysis enables policymakers to target interventions where they matter most.

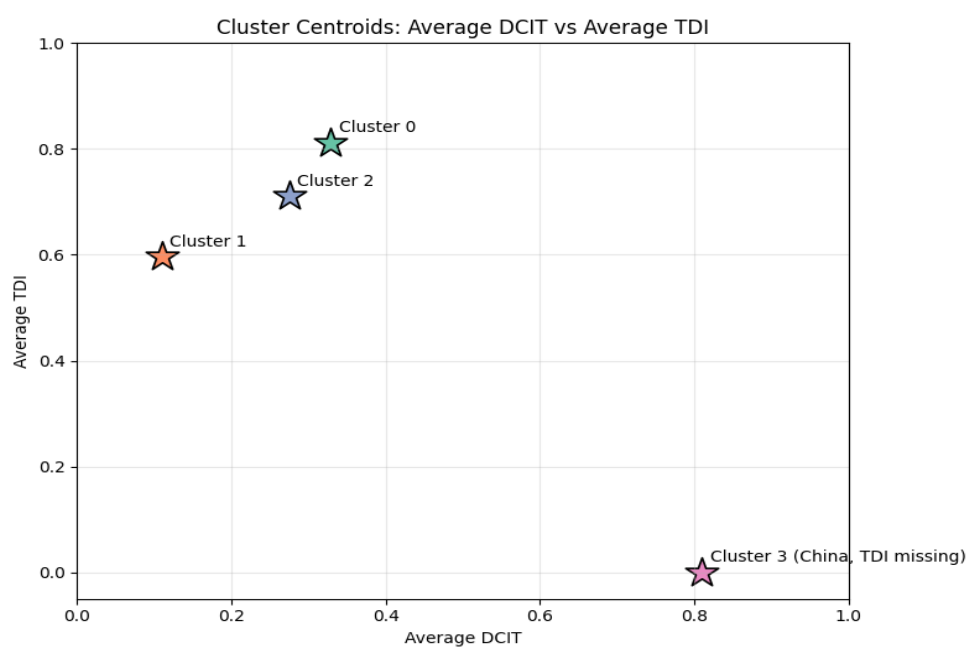

*Figure 6: Cluster Centroids*

Strategically, this comparative validation underscores the need for integrated approaches. Readiness alone is insufficient without governance and legal interoperability, while procedural reforms cannot sustain momentum without a solid digital backbone. Policymakers should leverage DCTI for prioritizing infrastructure and investment sequencing, while using TDI as a complementary benchmark to target regulatory modernization and cross-border harmonization. Together, these metrics provide a dual lens for designing interventions that align capability with execution, ensuring that digital trade strategies deliver inclusive and scalable outcomes.

## 6. Policy Implications

### 6.1 Forecast driven policy alignment

The following table bridges forecasting insights with actionable strategies, ensuring that policy decisions are grounded in projected competitiveness trends. By comparing DCIT values for 2024 and 2028 under the High Growth scenario, the table highlights which clusters will experience the most significant gains and where interventions should be prioritized. This forward-looking approach enables policymakers to sequence investments, design synergy corridors, and allocate resources effectively to maximize digital trade competitiveness.

| Cluster | DCIT 2024 | DCIT 2028 (High Growth) | Key Policy Actions |
|---|---|---|---|
| Cluster 0 (Industrializing) | 0,327 | 0,95 | Upgrade ICT infrastructure, integrate advanced tech, incentivize private investment |
| Cluster 1 (Low-Income) | 0,116 | 0,88 | Expand affordable connectivity, seek international funding, build customs automation capacity |
| Cluster 2 (Middle-Income) | 0,274 | 0,991 | Scale ICT adoption for SMEs, attract innovation-driven FDI, strengthen regional integration |
| Cluster 3 (Advanced/China) | 0,81 | 0,923 | Maintain R&D leadership, harmonize cross-border standards, expand digital services exports |

*Table 3: Cluster Forecast Values (2024 vs. 2028) and Scenario-Aligned Policy Actions*

By embedding forecast-based insights into policy design, decision-makers can move from reactive measures to proactive strategies, ensuring that investments and reforms are timed to maximize competitiveness gains projected for 2028 and beyond.

### 6.2 Cluster-Specific Policy Matrix

Table 4 (provided in appendix) presents cluster-specific priority actions compared to TDI:

1. Cluster 0 (Industrializing: Indonesia, Brazil): Upgrade ICT infrastructure (broadband, cloud), integrate advanced tech (AI, blockchain), incentivize private investment in digital logistics.

2. Cluster 1 (Low-Income: Nigeria, Myanmar): Expand affordable connectivity, seek international funding for infrastructure, build capacity in customs automation, promote mobile-first trade solutions.

3. Cluster 2 (Middle-Income: South Africa, Turkey, Argentina, Egypt): Scale ICT adoption for SMEs, attract innovation-driven FDI, strengthen regional digital integration, invest in cybersecurity and data governance.

4. Cluster 3 (China): Maintain leadership via R&D (AI, IoT, blockchain), harmonize cross-border digital standards, expand digital services exports, improve transparency in reporting.

## 6.3 Strategic Partnerships and ICT-Driven Trade Growth

Digitalization lowers trade costs, expands market access, and enables more inclusive participation, especially for Micro, Small, and Medium Enterprises (MSMEs). Cross-border digital partnerships accelerate trade and reduce transaction costs. However, global implementation of paperless trade reaches ~69%, while cross-border paperless averages ~46%, indicating that legal interoperability and data-exchange capacity remain bottlenecks (World Bank, 2023).

### 6.3.1 ICT-driven corridors and capital flows

- Cluster 3 (China) serves as the technology anchor, providing ICT capabilities and capital to all other clusters. Its leadership role is essential for accelerating infrastructure deployment and innovation diffusion.

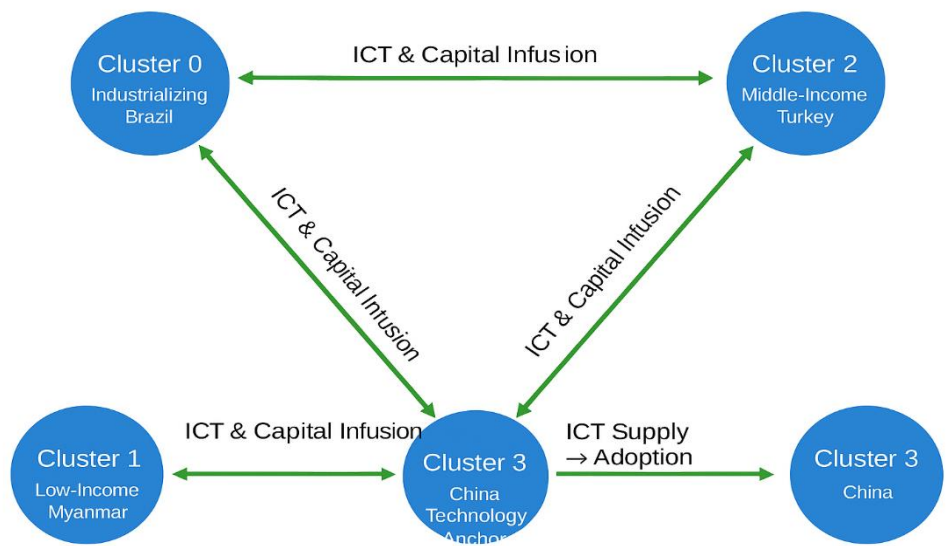

*Figure 7: Strategic Partnerships for ICT-Driven Trade Growth*

- Clusters 0 (Industrializing) and 2 (Middle-Income) form a mutual ICT supply–adoption corridor, enabling regional integration and shared standards for digital trade.
- Cluster 1 (Low-Income) relies heavily on capital infusion and ICT transfer from advanced clusters to overcome affordability and infrastructure gaps.

**Strategic Implications:**

- Prioritize synergy corridors: Pair advanced clusters with industrializing and middle-income economies to accelerate technology transfer and investment.
- Sequence interventions: Deploy ICT infrastructure first, then leverage FDI to scale and interconnect systems.
- Embed cooperation in regional frameworks: Align corridors with AfCFTA, ASEAN, and MERCOSUR for harmonized standards and cross-border paperless trade.
- Empower MSMEs: Ensure corridors include programs for SME digital adoption to maximize inclusive growth.

### 6.3.2 Actionable Recommendations

1. Establish a Digital Trade Competitiveness Platform integrating DCTI, TDI, and country-specific data to enable real-time benchmarking and strategic monitoring.

2. Co-design ICT and FDI investment packages targeting Cluster 1 and 2 countries, leveraging concessional finance, technical assistance, and South-South cooperation.

3. Harmonize cross-border data governance and digital standards across clusters via regional frameworks, reducing interoperability barriers and building trust in digital flows.

4. Develop targeted capacity-building programs for SMEs in mid-tier clusters to accelerate digital adoption, focusing on e-commerce, AI readiness, and cyber resilience.

5. Establish Regional Digital Trade Hubs in Clusters 0 and 2 to serve as technology anchors, knowledge centers, and platforms for regional integration.

6. Embed DCTI forecasts into national digital trade strategies, ensuring alignment of ICT and FDI policies with regional and global initiatives (African Continental Free Trade Area -AfCFTA, ASEAN (Association of Southeast Asian Nations -ASEAN).

### 7. Conclusion

This study introduces the Digital Competitiveness Index for Trade (DCIT), a robust, policy-responsive composite metric that captures enabling conditions for digital trade competitiveness across a group of emerging countries. Three key contributions emerged: (1) DCIT demonstrates high methodological robustness and strong policy responsiveness under High Growth scenario; (2) predictive analysis confirms DCIT as a powerful indicator of digital infrastructure and trade performance, though weaker for GDP growth and FDI, validating its trade-centric design; and (3) scenario-based forecasting reveals that ICT-first strategies and ICT+FDI synergy consistently outperform single-lever approaches, with synergistic benefits rising by cluster maturity (up to +10% additional gain).

The cluster-based analysis identifies differentiated pathways for digital trade acceleration: Cluster 3 (China) as a global technology leader; Clusters 0 and 2 as regional integrators and adopters with distinct roles in technology transfer and capacity building. Integration of DCITI with Trade Digitalization Index (TDI) benchmarking enables policymakers to align capability development with execution reforms, bridging the gap between readiness and implementation.

Data limitations, missing TDI values for Cluster 3 (China), limited time series for certain emerging markets, and reliance on proxy measures for digital trade, suggest future research directions: longitudinal extension to 2030, incorporation of real-time digital trade flow data, and subnational analysis to capture regional heterogeneity within large emerging economies. Nonetheless, DCIT provides governments and international organizations with a forward-looking, actionable framework for benchmarking digital trade readiness and prioritizing integrated ICT and FDI investment strategies aligned with inclusive growth and sustainable development objectives.

# Appendix

| Cluster | Current Profile | Priority Actions |
|---|---|---|
| Cluster 0 Industrializing | Moderate DCIT (≈0.33), High TDI (≈0.81) | - Upgrade ICT infrastructure (broadband, cloud)<br>- Integrate advanced tech (AI, blockchain)<br>- Incentivize private investment in digital logistics |
| Cluster 1 Low-Income | Low DCIT (≈0.11), Mid TDI (≈0.60) | - Expand affordable connectivity<br>- Seek international funding for infrastructure<br>- Build capacity in customs automation<br>- Promote mobile-first trade solutions |
| Cluster 2 Middle-Income | Balanced DCIT (≈0.28), TDI (≈0.71) | - Scale ICT adoption for SMEs<br>- Attract innovation-driven FDI<br>- Strengthen regional digital integration<br>- Invest in cybersecurity and data governance |
| Cluster 3 China | High DCIT (0.81), TDI missing | - Maintain leadership via R&D (AI, IoT, blockchain)<br>- Harmonize cross-border digital standards<br>- Expand digital services exports<br>- Improve transparency in reporting |

*Table 4: Policy Matrix by Cluster*